# Normalized Contact Force to Minimize "Electrode–Lead" Resistance in a Nanodevice

**Seung-Hoon Lee[1], Jun Bae[2], Seung Woo Lee[2], and Jae-Won Jang [1,*]**

[1]Department of Physics, Pukyong National University, Busan 608-737, KOREA

[2]School of Chemical Engineering, Yeungnam University, Gyeongsan, 712-749, KOREA

In this report, the contact resistance between "electrode" and "lead" is investigated for reasonable measurements of samples' resistance in a polypyrrole (PPy) nanowire device. The sample's resistance, including "electrode–lead" contact resistance, shows a decrease as force applied to the interface increases. Moreover, the sample's resistance becomes reasonably similar to, or lower than, values calculated by resistivity of PPy reported in previous studies. The decrease of electrode–lead contact resistance by increasing the applying force was analyzed by using Holm theory: the general equation of relation between contact resistance ($R_H$) of two-metal thin films and contact force ( $R_H \propto 1/\sqrt{F}$ ). The present investigation can guide a reliable way to minimize electrode–lead contact resistance for reasonable characterization of nanomaterials in a microelectrode device.

 *Email: jjang@pknu.ac.kr

Fax: +82-51-629-5575

## I. INTRODUCTION

In order to have efficient performance of nanodevices, the junction between microelectrode and sample nanomaterials (as well as properties of nanomaterials) is important. A good junction, such as "Ohmic contact", is favorable for high quality nanomaterials-based applications. In real applications, nanodevices should be connected by electrical leads into apparatus that monitors performance of the nanodevice. Besides the nanomaterials–electrode junction, good an "electrode–lead" junction is necessary for optimal performance of nanodevices. Efforts to achieve reliable junctions in nanodevices can be considered as obtaining lower metal–metal or metal–semiconductor contact resistance. It has been reported that the contributions to contact resistance include micro-scale geometry [1-5], oxide layer of semiconductor materials [6,7], difference of Fermi energy between nanomaterials and electrodes [8], and contact force between two metal films [3]. Furthermore, dependence on contact force [3,9] was investigated, as well as dependence on contact length [10], to understand characteristics of contact resistance. It seems that contact resistance can be altered by many elements, and not all of these are easy to be controlled. Challenges to minimize contact resistance have been investigated by several methods, including ultra-low contact resistance by epitaxial interfacing of nanowires to electrodes [11] and pre-process to remove oxide layer before deposition of nanomaterials on electrodes [6,7].

Conventional photolithography and/or e-beam lithography (EBL) are generally used to make microelectrodes in nanodevices [12-14]. Microelectrodes are integrated into apparatus to supply signals and monitor response of nanomaterials. Usually, conductive pastes are used to connect between microelectrodes and electrical leads of apparatus [15,16]. However, lower electrical conductivity, poor impact strength, and decreasing conductivity by humidity aging or normal use condition have been reported in reliability testing using silver paste as compared with tin-lead solder.[17] As it is difficult to use tin-lead soldering in microelectrodes for lead connections, more efforts to develop electrode–lead junctions better than silver pasting are needed. In addition, although the electrode–lead contact resistance caused by connection of microelectrodes to apparatus is also an empirically important factor, there have been few reports on "electrode-lead" contact resistance.

In the present study, resistance of nanodevices, composed of polypyrrole (PPy) nanowires (NWs) loaded on Au microelectrodes, was measured with non-silver paste electrode–lead junction in room conditions. Ag wires were physically contacted on the Au microelectrodes to transfer electrical signals between PPy NWs and an electrometer, and force was applied to the normal direction of the wire-electrode contact. The force dependent resistance of the device showed that electrode–lead contact resistance can be effectively minimized by increase of the applying force in accordance with Holm theory; this demonstration can be utilized for reasonable characterization of nanomaterials in microelectrode devices.

## II. EXPERIMENTS AND DISCUSSION

The experimental set-up for the force dependent electrode–lead contact resistance is displayed in Figure 1. PPy NWs were fabricated by electrochemical deposition into aluminum oxide (AAO) template using $ClO_4$ as a count ion. The PPy NWs were dispersed on a piece of Au microelectrodes patterned $SiO_x$/Si substrate (1.25 mm × 12.5 mm × 0.55 mm), as shown in the inset image in Figure 1. For the electrode–lead junction, pure Ag wires (99.99%, GoodFellow, Inc.) were put on the Au microelectrodes without any conductive paste or adhesive. Instead, the Ag wires physically contact onto the microelectrodes sandwiched between screws connected to two acrylic plates. Forces applying to the "Au electrode–Ag lead" junction can be adjusted by rotations of the screws, which are equally rotated [Figure 1(b)]. Measurement of the applying force was carried out by a force sensor (Cl-6537, Pasco scientific, Inc; (see *Supporting Information*). As represented in Figure 1(c), the applying force shows a linear dependence on the distance between the acrylic plates [H and $\Delta H$ in Figure 1(b): $\Delta H$ means displacement of H]. The force changes linearly from 2.45 N to 5.7 N as the H value decreases from 3.825 to 3.425 mm. From this linear dependence, applying forces during experiments were gauged and extrapolated by the H values with the screws rotating. Resistance of PPy NW devices were measured with a source-meter (2400, Keithley Instrument, Inc). For a control experiment, resistance of PPy NW device with a silver paste (Dotite D-500) contact Au electrode–Ag lead junction was also measured. $O_2$ plasma treatments of Au microelectrodes were carried out by a plasma cleaner (Harrick Plasma), with 11W for 10 min. All PPy NW devices were prepared with the same configuration (only one strand loading, equal length of PPy NW bridging between the microelectrodes) to minimize resistance deviation caused by PPy NW in different devices.

According to configuration of the PPy NW device shown in Figure 1(a). Total resistance of the device ($R_T$) will be straightforwardly represented by Eq. (1):

$$R_T = R_C + R_{PPy}, \qquad (1)$$

where $R_{PPy}$ is resistance of PPy NW and $R_C$ is contact resistance originated from nanomaterials–electrode contact and/or electrode–lead contact. In the case of a PPy NW device put on Au microelectrodes, Ohmic I-V curves were reported [18,19]. Therefore, $R_C$ can be regarded as mainly contributed by metal–metal junction (Au electrode–Ag wire lead), not by semiconductor-metal junction (PPy NW–Au microelectrode; Figure S2). The $R_T$ was measured with different forces that are applying to the normal direction of "Au electrode–Ag lead" junction (electrode–lead contact resistance *vs.* contact force). Figure 2(a) shows a decrease of $R_T$ as the contact force between the Au electrode–Ag lead" junction ($F$) increases; the *x*-axis of Figure 2(a) is normalized by the maximum contact force ($F_{Max}$) to the junction without deformation of the acrylic plates. Decrease of $R_T$ seems saturated in relatively strong contact force region ($F/F_{Max} \sim 0.8$). The dotted line of "a" in Figure 2(a) is resistance of a PPy NW device with Au electrode–Ag lead junction using silver paste. In addition, resistance of the PPy NW device calculated by using resistivity of PPy NW (undoped) [13] is displayed as the dashed line of "b" in Figure 2(a). In Figure 2(b), $R_T$ obtained by different cases are compared using a bar graph, where $R_T$ were measured with or without $O_2$ plasma treatment on the microelectrodes before making junction [denoted by $F_{Max}$ and $F_{Max}$ ($O_2$ plasma), respectively]. "Ag paste" and "PPy NW" in Figure 2(b) represent "a" and "b" of figure 2(a), respectively, while "PPy film" is resistance of a PPy NW device calculated by using resistivity of PPy film ($ClO_4$ as a count ion) reported in the literature [20]. Figure 2(b) shows that the "Ag paste" is larger than the "PPy film" and "PPy NW," which means the Au electrode–Ag lead junction using silver paste does not effectively reduce the junction resistance. On the other hand, the $F_{Max}$ and the $F_{Max}$ ($O_2$ plasma) are smaller than the "PPy film" and "PPy NW," which indicates that resistivity of our PPy NWs ($ClO_4$ as a count ion) is superior to that of PPy film ($ClO_4$ as a count ion) and PPy NW (undoped). This shows that the junction resistance is effectively reduced by contact force applying without using silver paste. In addition, the junction resistance can be reduced more by means of $O_2$ plasma treatment of the Au microelectrodes.

The contact force dependence of $R_T$ in Figure 2(a) can be explained by traditional contact theory in two-flat-metal-thin films, so it called "Holm theory." In general, contact resistance in Holm theory, $R_H$, can be described by Eq. (2) [3]:

$$R_H = \frac{\rho_1}{4a} + \frac{\rho_2}{4a}, \qquad (2)$$

where $a$ is the contact spot radius, $\rho_1$ and $\rho_2$ are electrical resistivity of each metal film (regard the metal films as metal 1 and metal 2, respectively). As the contact spot between two films becomes larger by increasing perpendicularly-applied contact force between the two films, the contact spot radius with applying force – that is Holm radius ($a_H$) – can be represented by the contact force ($F$) and hardness ($H$) of material with holding the contact force [3]:

$$a_H = \sqrt{\frac{F}{\pi H}}. \qquad (3)$$

From Eqs. 2 and 3, the relation between the contact force and the contact resistance $R_H$ can be obtained by Eq. (4):

$$R_H \propto \frac{1}{\sqrt{F}}. \tag{4}$$

The Au electrode–Ag lead junction can be approximated as a junction between two-flat-metal-thin films; contact spot between Ag wire and Au electrode is changed by the contact force between the Au electrode–Ag lead junction ($F$). The contact resistance, $R_C$, of the Au electrode–Ag lead junction can be systematically measured by putting two identical Ag wires on an electrically connected Au electrode.(Figure S3) The contact force dependence of $R_C$ can be analyzed with normalized contact force $F_N$:

$$R_C \propto \frac{1}{\sqrt{\dfrac{F - F_{Min}}{F_{Max}}}} = \frac{1}{\sqrt{F_N}}, \tag{5}$$

where $F_{Min}$ is the minimum contact force during the measurement; resistance cannot be measured by the electrometer if too weak a force is applied to the Au electrode–Ag lead junction. Figure 3(a) represents that normalized $R_C$ linearly increases with increase of $1/\sqrt{F_N}$ as expected in Eq. (5). The normalized $R_C$ was obtained by dividing by the minimum contact resistance; the minimum contact resistance was determined by the stable contact resistance with the contact force larger than $F_{Max}$ applied. Due to the relation with $F_N$ [shown in Figure 3(a)], the $R_C$ can be qualitatively regarded as contact resistance governed by Holm theory.

Moreover, the $R_T$ also shows linear dependence on $1/\sqrt{F_N}$, as shown in Figure 3(b). Because the total resistance $R_T$ is defined as Eq. (1), the contact resistance $R_C$ will be more dominant than $R_{PPy}$ (resistance of PPy NW) in the PPy NW device during the experiment. If $R_C$ is efficiently reduced, $R_T$ will be dominated by $R_{PPy}$. Such an efficient reduction of $R_C$ was accomplished by applying contact force to the junction as shown in Figure 2(b); $R_T$ shows the even lower resistance than reported values. According to force dependence of $R_T$ curve shown in Figure 2(a), >80% of the maximum force is necessary to efficiently reduce contact resistance in the PPy NW ($ClO_4$ as count ion) device. We successfully demonstrated that electrical properties of nanodevice can be reasonably measured by efficient contact force without silver paste soldering.

## III. CONCLUSION

This letter investigates the total resistance $R_T$ of a nanodevice as well as the contact resistance $R_C$ of electrode–lead junction behavior like contact resistance ruled by Holm theory. Based upon this observation, it was successfully demonstrated to minimize the experimental error of contact resistance originated from electrode–lead junction. In particular, the measured resistance with at least 80% of the maximum applying force to the junction without deformation of the apparatus shows reasonable values without experimental error such as riskily high contact resistance. This investigation can suggest a reasonable characterization of nanomaterials in a microelectrode device to minimize electrode–lead contact resistance.

## ACKNOWLEDGEMENT

J.-W. Jang appreciates for supporting by Basic Science Research Program through the National Research Foundation of Korea (NRF) funded by the Ministry of Education (NRF-2013K1A3A1A32035429, NRF-2013K2A2A4003362).

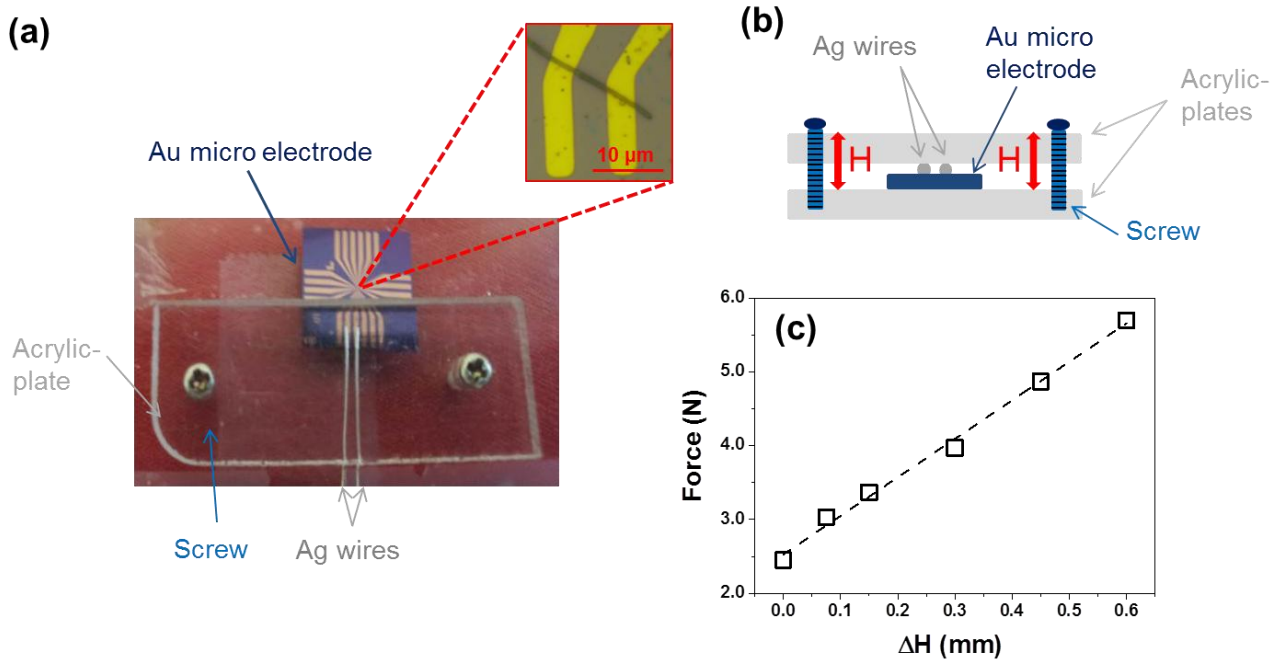

Fig. 1. (a) A snap-shot image and (b) scheme of experimental set-up for contact force dependent resistance measurement of a PPy NW device. The inset in (a) is a zoomed-in optical microscope image of PPy NW loaded area. (c) Applying force curve dependent on displacement (ΔH) of the distance between the acrylic plates (H).

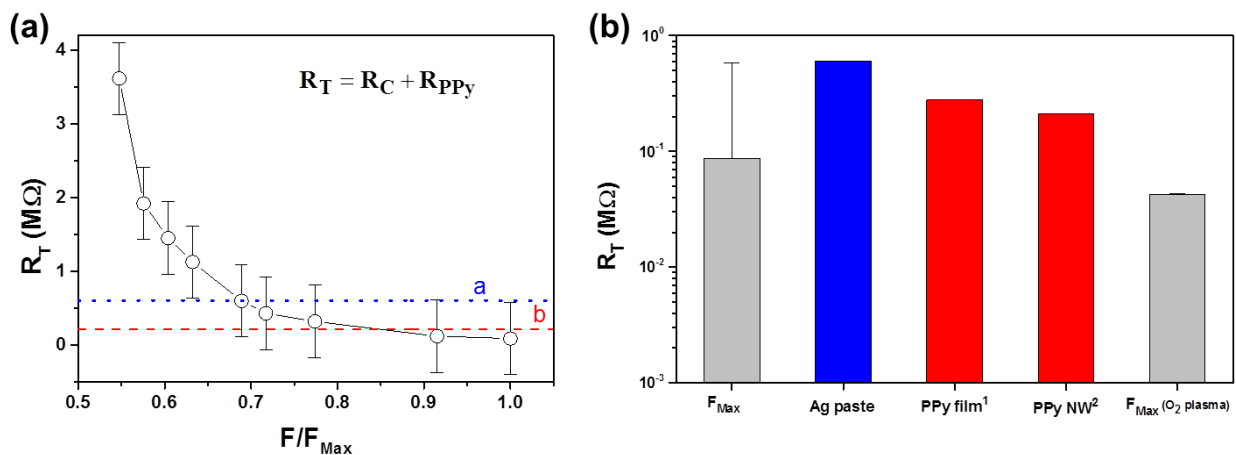

Fig. 2. (a) *Electrode–lead contact resistance vs. contact force*: total resistance of the device ($R_T$) dependent on contact force normalized by the maximum contact force ($F/F_{Max}$). "a" is resistance of a PPy NW device with Au electrode–Ag lead junction using silver paste, and "b" is resistance of the PPy NW device calculated by using resistivity of PPy NW (undoped). $R_T$ is represented by the sum of contact resistance ($R_C$) in the device and resistance of PPy NW ($R_{PPy}$). (b) The bar graphs of $R_T$ with the maximum contact force ($F_{Max}$), junctioned by Ag paste, and the maximum contact force sequentially after $O_2$ plasma treatment [$F_{Max}$($O_2$ plasma)]; and the bar graphs of resistances of a PPy NW device using resistivity of PPy film (PPy film[1]) and PPy NW (PPy NW[2]). [1]Resistivity of PPy film ($ClO_4$ as a count ion)[20]; [2]Resistivity of PPy NW (undoped) [13].

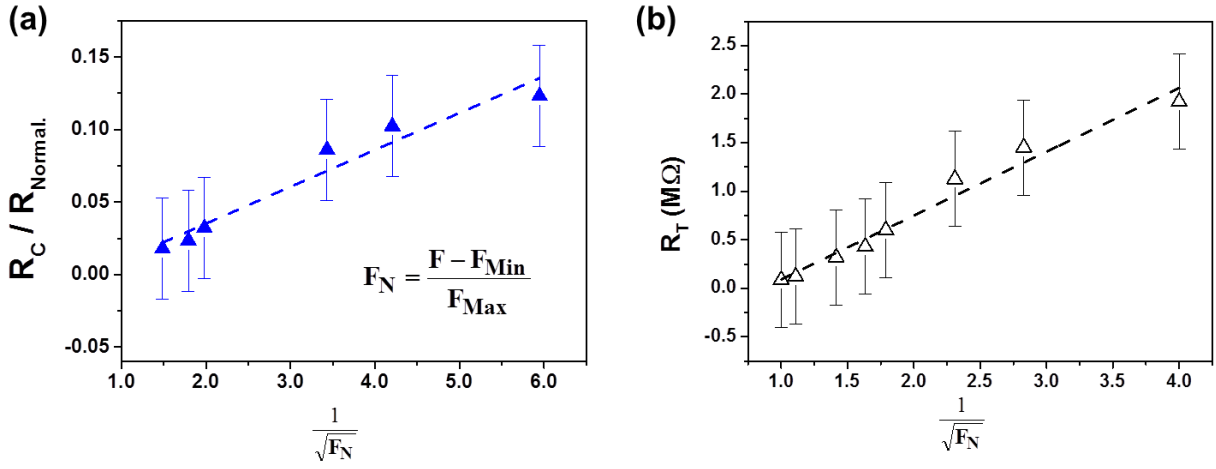

Fig. 3. Graphs of (a) normalized $R_C$ and (b) $R_T$ dependent on $1/\sqrt{F_N}$; both show a linear correlation. The normalized applying force ($F_N$) is determined by the equation in (a): the $F_{Max}$ and $F_{Min}$ are the maximum and minimum contact force during the measurement, respectively.

Supporting Information

# Normalized Contact Force to Minimize "Electrode–Lead" Resistance in a Nanodevice

**Seung-Hoon Lee[1], Jun Bae[2], Seung Woo Lee[2], and Jae-Won Jang [1,*]**

[1]*Department of Physics, Pukyong National University, Busan 608-737, KOREA*

[2]*School of Chemical Engineering, Yeungnam University, Gyeongsan, 712-749, KOREA*

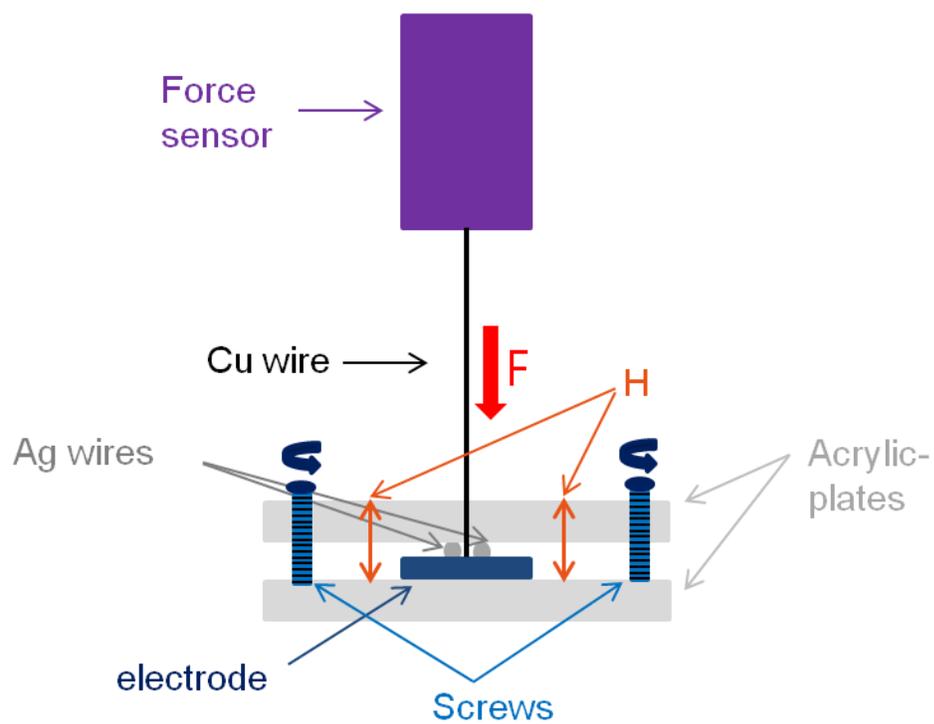

Fig. S1. Scheme of experimental set-up for contact force (*F*) dependent on the distance between the acrylic plates (H).

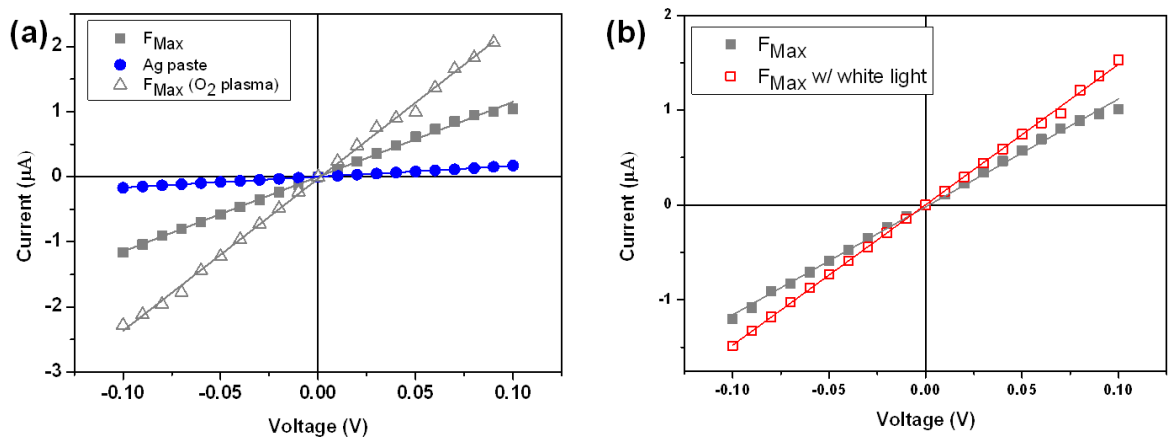

Fig. S2. (a) *Voltage–Current* curves of the PPy NW device with the maximum contact force ($F_{Max}$), junctioned by Ag paste, and the maximum contact force sequentially after O$_2$ plasma treatment [$F_{Max}$ (O$_2$ plasma)]. Ohmic behavior of curves (linear I-V) can be regarded as indicating that contact resistance between the PPy NW and the Au microelectrode is negligible. (b) *Voltage – Current* curves of the PPy NW device with the maximum contact force: Dark current measurement ($F_{Max}$) and photocurrent measurement ($F_{Max}$ w/ white light) by a halogen lamp (~10W) irradiation. The increased photocurrent reflects that the electrical property of the device with $F_{Max}$ is dominated by photo-response of PPy NW.

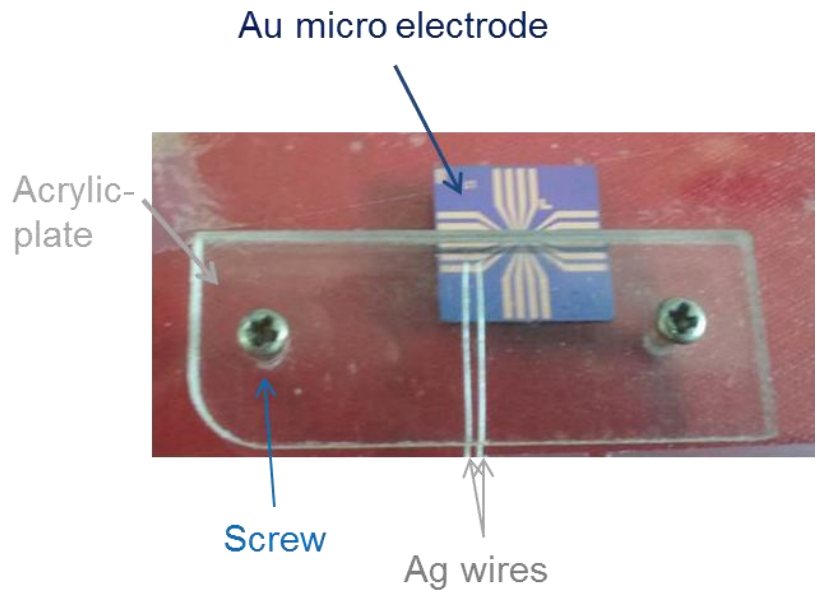

Fig. S3. A snap-shot image of experimental set-up for contact force-dependent contact resistance ($R_C$) measurement. The Ag wires are connected on the same Au electrode.